\documentclass[aps,twocolumn,showpacs]{revtex4}%
\usepackage{amsfonts}
\usepackage{amsmath}
\usepackage{amssymb}
\usepackage{appendix}
\usepackage{graphicx}%
\providecommand{\U}[1]{\protect\rule{.1in}{.1in}}
\providecommand{\U}[1]{\protect\rule{.1in}{.1in}}

\begin{document}
\title{Optomechanically induced ultraslow and ultrafast light}
\author{\ Xiao-Bo Yan}
\email{xiaoboyan@126.com}
\affiliation{College of Electronic Science, Northeast Petroleum University, Daqing  163318, China}
\date{\today}


\begin{abstract}
Slow and fast light is an important and fascinating phenomenon in quantum optics. Here, we theoretically study how to achieve the ultraslow and ultrafast light in a passive-active optomechanical system, based on the ideal optomechanically induced transparency (OMIT). Under the conditions of the ideal OMIT, an abnormal (inverted) transparency window will emerge accompanied with a very steep dispersion, resulting that the ultraslow light can be easily achieved at the transparency window by adjusting the dissipation rates of the two cavities, even with usual mechanical linewidth (such as Hz linewidth). Particularly, as the decay rate of the passive cavity tends to the gain rate of the active cavity, the ideal stopped light can be achieved. Similarly, the ultrafast light can be achieved at transparency window by tuning the coupling strength and the decay rates in the system.
\end{abstract}
\maketitle

\section{Introduction}

Over the last few decades, slow and fast light has fascinated the optical
physics community for the exotic phenomenon where the group velocity of light can be larger (even be negative) or less than the speed of light in a vacuum, and for the appealing applications in optical telecommunication  and interferometry \cite{Boyd2009sci,Zimmer2004prl,Shahriar2007pra,Shi2007ol}. 
In general, it is critical to obtain slow and fast light that the material system can exhibit an abnormal dispersion when the pulse propagates through. Fortunately, the technology based on ``Electromagnetically
Induced Transparency" (EIT) \cite{Harris1997phytoday} exactly can provide the platform. At the transparency window accompanied with an abnormal dispersion, the group velocity of light passing through the material will change dramatically. This aspect of the effect has been utilized to conjure schemes whereby light may turn slow or fast \cite{Kash1999prl,Budker1999prl,Hau1999Nat,Wang2000Nat,Turukhin2001prl,Yanik2004prl,Totsuka2007prl,Zhang2009prl,Klein2009pra}. Such as, taking advantage of EIT in ultracold atom gases, light could be
slowed down to the ``human" scale of 17 m/s \cite{Hau1999Nat} and light storage has been demonstrated in Bose-Einstein Condensates \cite{Zhang2009prl}.

In parallel, cavity optomechanics \cite{Aspelmeyer2014} exploring the interaction between engineered optical and mechanical modes has been sufficiently studied, where various interesting quantum phenomena can be realized, such as mechanical ground-state cooling
\cite{Marquardt2007,Wilson-Rae2007,BingHe2017}, quantum entanglement \cite{Deng2016,Vitali2007,Yan2019OE}, photon antibunching
\cite{Rabl2011prl,Xu2013pra,Wang2019pra,Liao2013pra}, especially optomechanically induced transparency (OMIT) \cite{Karuza2013,Huang2010_041803,Weis2010,Yan2015,Safavi-Naeini2011,Huang2011,Jing2015,Li2016sr,LiuYC2017,Yan2020pra,Shahidani2013,Chen2011,LiuYX2013,Kronwald2013,Lu2017,Lu2018,Dong2013,Dong2015,Ma2014pra,Xiong2012,Xiong2018,Bodiya2019pra,Zhang2017pra} and the OMIT-based slow and fast light  \cite{Chen2011,Chang2011njp,Jing2015,Tarhan2013,Akram2015,Gu2015,Safavi-Naeini2011}. Nevertheless, all these studies of slow and fast light are not based on ideal OMIT. In fact, the conditions of the ideal OMIT are very critical for the response of optomechanical systems to the signal light \cite{Yan2020pra}.

Recently, we studied the slow light effect in one-cavity optomechanical system \cite{Yan2020slow} based on the conditions of ideal OMIT according to the method in Ref. \cite{Yan2020pra}. We find there is only slow light effect in the one-cavity optomechanical system, and the upper bound of the optical group delay of the transmitted probe light is exactly equal to the mechanical ringdown time. It means that if we want to obtain significant slow light effect, we have to manufacture the mechanical resonator with small enough dissipation rate. However, this is not a easy thing to do under the current experimental conditions, although now the mechanical oscillators with dissipation rate of millihertz can be manufactured experimentally \cite{Norte2016prl,Reinhardt2016prx,Ghadimi2018sci,Tsaturyan2017NatN}. Therefore, we want to know whether there exist some other proposals whereby the significant optical group delay can easily exceed the upper bound (mechanical ringdown time) even with usual mechanical bandwidth, and whether the fast light can be realized at the transparency window of the ideal OMIT.

Here, we theoretically study how to realize ultraslow and ultrafast light based on the conditions of the ideal OMIT in a passive-active optomechanical system (a passive optomechanical cavity coupled to an active cavity, see Fig. 1) \cite{Zhang2017pra,Jing2015,Li2016sr}.
The existence of the active cavity can strongly affect the response of the system to the probe field.
First, we obtain the conditions of the ideal OMIT in the system according to the method in Ref. \cite{Yan2020pra}. Under these conditions, we find an abnormal (inverted) transparency window will emerge, accompanied with a very steep and negative dispersion. Secondly, we obtain the expression of the optical group delay and that of the dispersion slope at the transparency window, and find the optical group delay is exactly equal to the negative value of the dispersion slope. Thirdly, the ultraslow light can be easily achieved at the transparency window even with the usual mechanical linewidth (such as Hz linewidth), particularly when the decay rate of the passive cavity tends to the gain rate of the active cavity, the ideal stopped light can be achieved. Finally, not like the case in one-cavity optomechanical system where there is only slow light effect, in this system, the ultrafast light can be easily achieved by adjusting the coupling strength between the two cavities and the decay rates.  We believe the results can be used to control optical transmission in modern optical networks.

\section{Model and equations}

\begin{figure}[ptb]
	\includegraphics[width=0.42\textwidth]{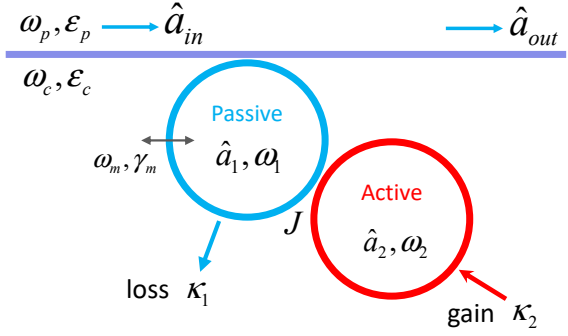}\caption{ Sketch of a passive-active optomechanical system. The cavity $\hat{a}_{1}$ is driven by a coupling field with frequency $\omega_{c}$ (amplitude $\varepsilon_{c}$) and a weak probe field with frequency $\omega_{p}$ (amplitude $\varepsilon_{p}$). The two cavities $\hat{a}_{1}$ and $\hat{a}_{2}$ are coupled to each other by interaction strength $J$.}%
	\label{Fig1}%
\end{figure}

We consider a system of two coupled
whispering-gallery-mode microtoroid resonators (see Fig. 1). The first resonator with frequency $\omega_{1}$ is passive and contains a mechanical mode with frequency $\omega_{m}$ and mass $m$. The second resonator with frequency $\omega_{2}$ is active and coupled to the first one through an evanescent
field. The coupling strength $J$ between the two resonators can be tuned by
changing the distance between them. The annihilation operator of the passive (active) cavity is denoted by $\hat{a}_{1}$ ($\hat{a}_{2}$), and the position and momentum operators of the mechanical resonator
are represented by $\hat{x}$ and $\hat{p}$, respectively. Cavity $\hat{a}_{1}$ is also coupled to the mechanical resonator via optomechanical interaction $-\hbar g_{0}\hat{a}_{1}^{\dagger}\hat{a}_{1}\hat{x}$ with the interaction strength $g_{0}=\omega_{1}/R$ and $R$ is the radius of the passive microtoroid cavity. The cavity $\hat{a}_{1}$ is driven by a strong coupling field with
frequency $\omega_{c}$ (amplitude $\varepsilon_{c}$) and a weak probe field with frequency $\omega_{p}$ (amplitude $\varepsilon_{p}$). Then, the Hamiltonian in the rotating frame at the frequency of the coupling field $\omega_{c}$ is 
\begin{eqnarray}
H&=&\hbar\Delta_{1}\hat{a}_{1}^{\dagger }\hat{a}_{1}+\hbar\Delta_{2}\hat{a}_{2}^{\dagger }\hat{a}_{2}+\frac{\hat{p}^{2}}{2m}+\frac{1}{2}m\omega_{m}^{2}\hat{x}^{2}\notag\\
&-&\hbar g_{0}\hat{a}_{1}^{\dagger}\hat{a}_{1}\hat{x}-\hbar J(\hat{a}_{1}^{\dagger}\hat{a}_{2}+\hat{a}_{2}^{\dagger}\hat{a}_{1})\notag\\
&+&i\hbar\varepsilon_{c}(\hat{a}_{1}^{\dagger}-\hat{a}_{1})
+i\hbar(\hat{a}_{1}^{\dagger}\varepsilon_{p}e^{-i\delta t}-\hat{a}_{1}\varepsilon_{p}^{\ast}e^{i\delta t}).  
\end{eqnarray}%
Here, $\delta=\omega_{p}-\omega_{c}$ ($\Delta_{1,2}=\omega_{1,2}-\omega_{c}$) is the detuning between the probe field (cavity field) and coupling field.

In this paper, we deal with the mean response of the system to the probe field in the presence of the coupling field, hence we do not include quantum fluctuations. 
Using the factorization assumption $\langle \hat{a}_{1}\hat{x}\rangle=\langle\hat{a}_{1}\rangle\langle\hat{x}\rangle$, and then the mean value equations can be given by
\begin{eqnarray}
\dot{a}_{1}&=&-(\kappa_{1}+i\Delta_{1}-ig_{0}x)a_{1}+iJa_{2}+\varepsilon_{c}+\varepsilon_{p}e^{-i\delta t},\notag\\
\dot{a}_{2}&=&(\kappa_{2}-i\Delta_{2})a_{2}+iJa_{1},\notag\\
\dot{p}&=&-\gamma_{m}p-m\omega_{m}^{2}x+\hbar g_{0}|a_{1}|^{2}.\notag\\
\dot{x}&=&\frac{p}{m},
\end{eqnarray}
Here, $\gamma_{m}$ is the mechanical damping rate and $\kappa_{1}$ ($\kappa_{2}$) is the decay (gain) rate of the passive (active) cavity. In the absence of the probe field $\varepsilon_{p}$, from Eq. (2) the mean values in the steady state can be obtained respectively as
\begin{eqnarray}
a_{1s}&=&\frac{\varepsilon_{c}}{\kappa_{1}+i\Delta_{1}-ig_{0}x_{s}+\frac{J^{2}}{i\Delta_{2}-\kappa_{2}}},\notag\\
a_{2s}&=&\frac{iJa_{1s}}{i\Delta_{2}-\kappa_{2}},\notag\\
x_{s}&=&\frac{\hbar g_{0}|a_{1s}|^{2}}{m\omega_{m}^{2}},\notag\\
p_{s}&=&0.
\end{eqnarray}

In the presence of the probe field $\varepsilon_{p}$, we can expand each mean value as the sum of its steady state value and a small fluctuation around that value, i.e., $a_{i}=a_{is}+\delta a_{i}$, $x=x_{s}+\delta x$ and $p=p_{s}+\delta p$.
Substituting them into Eq. (2) and keeping only the
linear terms, we obtain the linearized Langevin equations
\begin{eqnarray}
\delta\dot{a}_{1}&=&-(\kappa_{1}+i\bar{\Delta}_{1})\delta a_{1}+ig_{0}a_{1s}\delta x+iJ\delta a_{2}+\varepsilon_{p}e^{-i\delta t},\notag\\
\delta\dot{a}_{2}&=&(\kappa_{2}-i\Delta_{2})\delta a_{2}+iJ\delta a_{1}.\notag\\
\delta\dot{p}&=&-\gamma_{m}\delta p-m\omega_{m}^{2}\delta x+\hbar g_{0}a_{1s}^{\ast}\delta a_{1}+\hbar g_{0}a_{1s}\delta a_{1}^{\ast},\notag\\
\delta\dot{x}&=&\frac{\delta p}{m},
\end{eqnarray}
here, $\bar{\Delta}_{1}=\Delta_{1}-g_{0}x_{s}$.
Using the usual method \cite{Huang2010_041803,Weis2010,Safavi-Naeini2011,Karuza2013}, we can solve Eq. (4) by writing the solution in the form
$\delta s=s_{+}e^{-i\delta t}+s_{-}e^{i\delta t}$ ($s=a_{1},a_{2},x,p$). Inserting them into Eq. (4) and comparing the coefficients of $e^{\pm i\delta t}$ on both sides of the equation, then
we obtain
\begin{eqnarray}
a_{1+}=\frac{\varepsilon_{p}}{\kappa_{1}-i(\delta-\bar{\Delta}_{1})+\mathrm{A}+\frac{\beta}{\frac{\delta^{2}-\omega_{m}^{2}+i\delta\gamma_{m}}{2i\omega_{m}}-\frac{\beta}{\kappa_{1}-i(\delta+\bar{\Delta}_{1})+\mathrm{B}}}},
\end{eqnarray}
here,
\begin{eqnarray}
\beta&=&\frac{\hbar g_{0}^{2}|a_{1s}|^{2}}{2m\omega_{m}},\\
\mathrm{A}&=&-\frac{J^{2}}{\kappa_{2}+i\delta-i\Delta_{2}},\\
\mathrm{B}&=&-\frac{J^{2}}{\kappa_{2}+i\delta+i\Delta_{2}}.
\end{eqnarray}
Based on Eq. (5), we can study the response of the optomechanical system to the probe field.
Because it is known that the coupling between the cavity and the resonator is strong at the near-resonant frequency, we consider $\bar{\Delta}_{1}=\Delta_{2}=\omega_{m}$ in this work.

\section{Ultraslow light at $\delta=\omega_{m}$}

The quadrature of the  optical components with frequency $\omega_{p}$ in the output field
can be defined as $\varepsilon_{T}=2\kappa_{1}a_{1+}/\varepsilon_{p}$ \cite{Huang2010_041803}. The real part $\mathrm{Re}[\varepsilon_{T}]$ and the imaginary part $\mathrm{Im}[\varepsilon_{T}]$ represent
the absorptive and dispersive behavior of the optomechanical system to the probe field, respectively. 
Since the slow and fast light is based on the properties of OMIT, we first give the conditions of ideal OMIT at $\delta=\omega_{m}$. According to the pole location of the subfraction in Eq. (5) \cite{Yan2020pra}, the conditions of ideal OMIT at $\delta=\omega_{m}$ can be obtained as
\begin{eqnarray}
\beta&=&\frac{\gamma_{m}(\kappa_{1}-\kappa_{2})}{2},\\
J&=&\pm\sqrt{\kappa_{2}^{2}+4\omega_{m}^{2}},
\end{eqnarray}
which means the ideal OMIT at $\delta=\omega_{m}$ can be achieved only when $\kappa_{1}>\kappa_{2}$ because $\beta>0$ according to Eq. (6).

\begin{figure}[ptb]
	\includegraphics[width=0.46\textwidth]{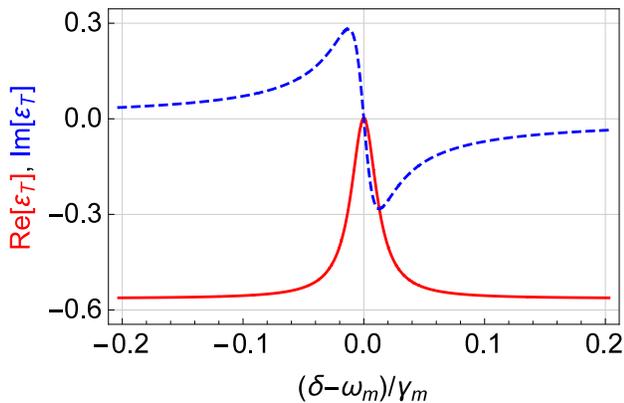}\caption{The real part $\mathrm{Re}[\varepsilon_{T}]$ (red-solid) and the imaginary part $\mathrm{Im}[\varepsilon_{T}]$ (blue-dashed) vs. normalized frequency detuning $(\delta-\omega_{m})/\gamma_{m}$ are plotted with parameters $\omega_{m}=\kappa_{1}/1.1=\kappa_{2}=10^{4}\gamma_{m}$ under the conditions in Eqs. (9) and (10).}%
	\label{Fig2}%
\end{figure}

In Fig. 2, we plot $\mathrm{Re}[\varepsilon_{T}]$ (red-solid) and $\mathrm{Im}[\varepsilon_{T}]$ (blue-dashed) vs. normalized frequency detuning $(\delta-\omega_{m})/\gamma_{m}$ according to Eq. (5) with parameters $\omega_{m}=\kappa_{1}/1.1=\kappa_{2}=10^{4}\gamma_{m}$ under the conditions in Eqs. (9) and (10).
It can be seen from Fig. 2 that with these parameters, the abnormal (inverted) transparency window emerges at detuning $\delta=\omega_{m}$ ($\mathrm{Re}[\varepsilon_{T}]<0$ near the transparency window), accompanied by a steep dispersion (blue-dashed) with a negative slope. In fact, the real part $\mathrm{Re}[\varepsilon_{T}]$ will become positive near the transparency window if $J^{2}<\kappa_{1}\kappa_{2}$ (we will not show the OMIT in the case because there is nothing else special, such as, the dispersion slope is also negative there).

\begin{figure}[ptb]
	\includegraphics[width=0.46\textwidth]{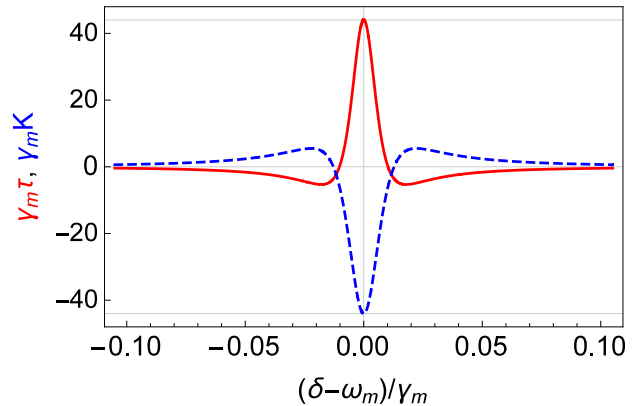}\caption{The normalized time delay $\gamma_{m}\tau$ (red-solid) and the dispersion curve slope
		$\gamma_{m}\mathrm{K}$ (blue-dashed) vs. normalized frequency detuning $(\delta-\omega_{m})/\gamma_{m}$ are plotted with the same parameters in Fig. 2.}%
	\label{Fig3}%
\end{figure}

In Fig. 3, we plot the normalized dispersion curve slope
$\gamma_{m}\mathrm{K}$ ($\mathrm{K}=\frac{\partial\mathrm{Im}[\varepsilon_{T}]}{\partial\omega_{p}}$ and see the blue-dashed line) vs. normalized frequency detuning $(\delta-\omega_{m})/\gamma_{m}$ with the same parameters in Fig. 2. It can be seen from Fig. 3 that the dispersion curve slope $\mathrm{K}$ will take the negative
maximum at the transparent window $\delta=\omega_{m}$. The expression of the slope $\mathrm{K}$ is too lengthy to be presented here. While it will become very simple at the transparent window $\delta=\omega_{m}$. According to Eq. (5), the slope at the transparency window can be obtained as
\begin{eqnarray}
\mathrm{K}_{\mathrm{max}}=-\frac{4\kappa_{1}[(\kappa_{1}-\kappa_{2})\kappa_{2}^{2}+4(\gamma_{m}+\kappa_{1}-\kappa_{2})\omega_{m}^{2}]}{\gamma_{m}(\kappa_{1}-\kappa_{2})^{2}(\kappa_{2}^{2}+4\omega_{m}^{2})}.
\end{eqnarray}
which means that the slope $\mathrm{K}_{\mathrm{max}}$ is always negative and will approach infinity in the limit of $\kappa_{1}\rightarrow\kappa_{2}$. We will see such steep dispersion behavior can cause the ultraslow light in the system.   

According to the input-output relation \cite{Aspelmeyer2014}, the
optical group delay of the transmitted light is given by \cite{Weis2010,Safavi-Naeini2011}
\begin{eqnarray}
\tau=\frac{\partial\mathrm{arg}[1-\varepsilon_{T}]}{\partial\omega_{p}}.
\end{eqnarray}
The positive (negative) value of the delay $\tau$ represents slow (fast) light \cite{Bigelow2003sci} in the system. 
In Fig. 3, we plot the normalized time delay $\gamma_{m}\tau$ (red-solid) vs. normalized frequency detuning $(\delta-\omega_{m})/\gamma_{m}$ with the same parameters in Fig. 2. It can be seen from Fig. 3 that the maximum time delay $\tau_{\mathrm{max}}$ occurs exactly at the transparent window.
According to Eqs. (5), (9), (10) and (12), we obtain the analytic expression of the maximum time delay $\tau_{\mathrm{max}}$ (at the transparent window $\delta=\omega_{m}$) as
\begin{eqnarray}
\tau_{\mathrm{max}}=\frac{4\kappa_{1}[(\kappa_{1}-\kappa_{2})\kappa_{2}^{2}+4(\gamma_{m}+\kappa_{1}-\kappa_{2})\omega_{m}^{2}]}{\gamma_{m}(\kappa_{1}-\kappa_{2})^{2}(\kappa_{2}^{2}+4\omega_{m}^{2})},
\end{eqnarray}
which is exactly equal to $-\mathrm{K}_{\mathrm{max}}$ in Eq. (11).
It means that the steeper the slope of dispersion curve is, the larger the slow light effect becomes. It can be seen from Eq. (13) that if $\kappa_{1}\rightarrow\kappa_{2}$, the group delay $\tau_{\mathrm{max}}$ will approach infinity, which means the light will be stopped.

It can be seen from Eq. (13) that if $\gamma_{m}\ll\kappa_{1}-\kappa_{2}$ resulting that the term $\gamma_{m}$ in the numerator can be ignored, then the $\tau_{\mathrm{max}}$ can be simplified as
\begin{eqnarray}
\tau_{\mathrm{max}}=\frac{4\kappa_{1}}{\gamma_{m}(\kappa_{1}-\kappa_{2})}.
\end{eqnarray}
It means in this case $\tau_{\mathrm{max}}$ will not be related to $\omega_{m}$, which can be verified in Fig. 4 where we plot the normalized time delay $\gamma_{m}\tau$ vs. the normalized detuning $(\delta-\omega_{m})/\gamma_{m}$ with $\omega_{m}=5\times10^{3}\gamma_{m}$ (blue-dashed), $\omega_{m}=10^{4}\gamma_{m}$ (black-dotted), $\omega_{m}=2\times10^{4}\gamma_{m}$ (red-solid), and $\kappa_{1}=2\kappa_{2}=2\times10^{4}\gamma_{m}$. It can be seen from Fig. 4 that the mechanical frequency $\omega_{m}$ does not affect the maximum value of the time delay, but will affect the width of the delay spectrum.

If $\kappa_{1}-\kappa_{2}\ll\gamma_{m}$ resulting that the term $\kappa_{1}-\kappa_{2}$ in the numerator in Eq. (13) can be ignored, the $\tau_{\mathrm{max}}$ can be simplified as
\begin{eqnarray}
\tau_{\mathrm{max}}=\frac{16\kappa_{1}\omega_{m}^{2}}{(\kappa_{1}-\kappa_{2})^{2}(\kappa_{2}^{2}+4\omega_{m}^{2})}.
\end{eqnarray}
It means that the time delay will not be related to $\gamma_{m}$, which can be verified in Fig. 5 where we plot the time delay $\tau_{\mathrm{max}}$ vs. the difference $\xi$ $(=\kappa_{1}-\kappa_{2})$ according to Eq. (13) with $\omega_{m}=\kappa_{2}=10^{4}$, $\gamma_{m}=10$ (red-line) and $\gamma_{m}=100$ (blue-dashed). It can be seen from Fig. 5 that the two curves with different $\gamma_{m}$ will tend to the same value $\tau_{\mathrm{max}}$ given by Eq. (15) as $\xi\ll\gamma_{m}$.      

From the above analysis, we can know the maximum time delay $\tau_{\mathrm{max}}$ in this system can easily exceed the mechanical ringdown time $\tau_{m}=\frac{2}{\gamma_{m}}$ \cite{Thompson2008} (the upper bound of time delay in one-cavity optomechanical system \cite{Yan2020slow}). Actually, the ultraslow light can be achieved in our present system even with the usual Hz linewidth. Such as, if $\gamma_{m}=0.76$ Hz \cite{Thompson2008} and $\kappa_{2}=\omega_{m}=10\xi=10^{4}$ Hz, the time delay $\tau_{\mathrm{max}}$ is about \textit{one minute} that is much larger than the mechanical ringdown time $\tau_{m}$ ($\simeq2.63$ second). If we adopt the mechanical resonator with mHz linewidth \cite{Norte2016prl,Reinhardt2016prx,Ghadimi2018sci,Tsaturyan2017NatN} and with the same other parameters, the time delay $\tau_{\mathrm{max}}$ can be enhanced to about half a day. These ultralong time delays
may be used for OMIT-based memories in the future.

\begin{figure}[ptb]
	\includegraphics[width=0.45\textwidth]{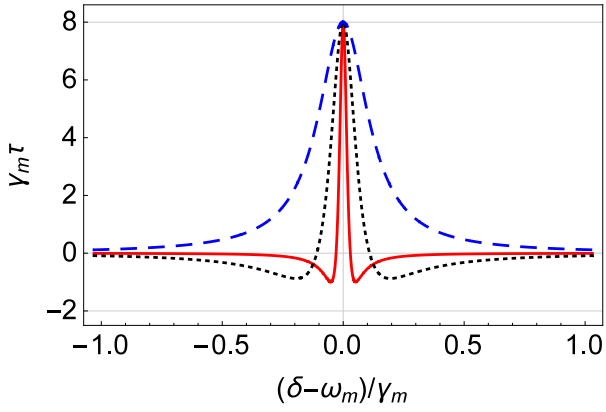}\caption{The normalized time delay $\gamma_{m}\tau$ vs. the normalized detuning $(\delta-\omega_{m})/\gamma_{m}$ is plotted with $\omega_{m}=5\times10^{3}\gamma_{m}$ (blue-dashed), $\omega_{m}=10^{4}\gamma_{m}$ (black-dotted), $\omega_{m}=2\times10^{4}\gamma_{m}$ (red-solid). Other parameters are $\kappa_{1}=2\kappa_{2}=2\times10^{4}\gamma_{m}$.}%
	\label{Fig4}%
\end{figure}

\begin{figure}[ptb]
	\includegraphics[width=0.46\textwidth]{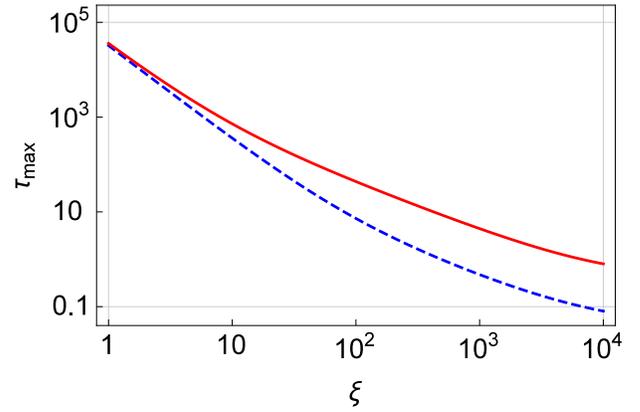}\caption{The maximum time delay $\tau_{\mathrm{max}}$ in Eq. (13) vs. the difference $\xi$ $(=\kappa_{1}-\kappa_{2})$ is plotted with parameters $\omega_{m}=\kappa_{2}=10^{4}$ and $\gamma_{m}=10$ (red-line), $\gamma_{m}=100$ (blue-dashed).}%
	\label{Fig5}%
\end{figure}

\section{Ultrafast light at $\delta=-\omega_{m}$}

\begin{figure}[b]
	\includegraphics[width=0.45\textwidth]{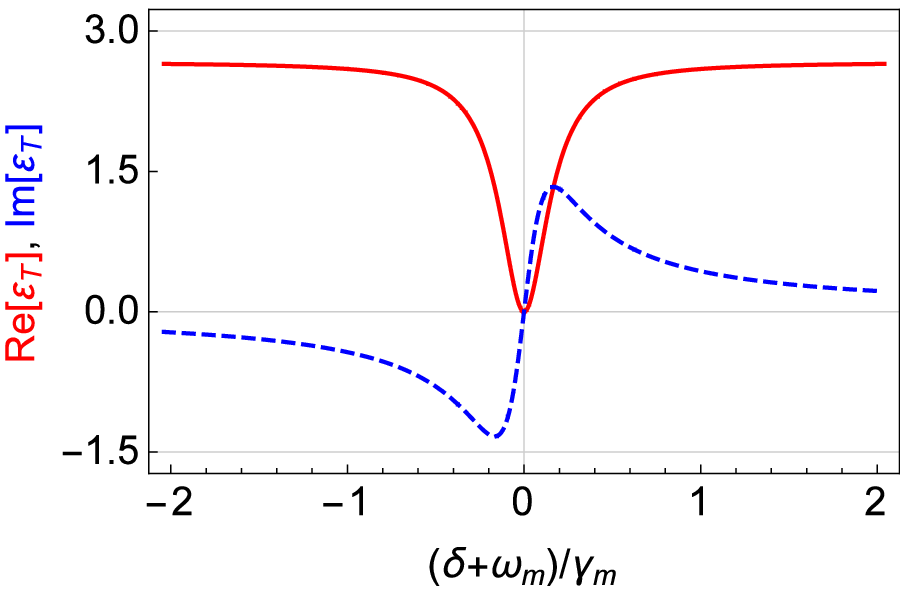}\caption{The real part $\mathrm{Re}[\varepsilon_{T}]$ (red-solid) and the imaginary part $\mathrm{Im}[\varepsilon_{T}]$ (blue-dashed) vs. normalized frequency detuning $(\delta+\omega_{m})/\gamma_{m}$ are plotted with parameters $\omega_{m}=\kappa_{1}/4=\kappa_{2}=10^{4}\gamma_{m}$ under the conditions in Eqs. (16) and (17).}%
	\label{Fig6}%
\end{figure}

Now we study the ultrafast light which can be achieved based on the OMIT at $\delta=-\omega_{m}$. According to Eq. (5), we find a beautiful transparency window at $\delta=-\omega_{m}$ will appear when
\begin{eqnarray}
\beta&=&\frac{\gamma_{m}(J^{2}-\kappa_{1}\kappa_{2})}{2\kappa_{2}},\\
J&=&\pm\sqrt{\kappa_{2}^{2}+4\omega_{m}^{2}},
\end{eqnarray}
which means that $J^{2}$ must be larger than $\kappa_{1}\kappa_{2}$.

In Fig. 6, we plot $\mathrm{Re}[\varepsilon_{T}]$ (red-solid) and $\mathrm{Im}[\varepsilon_{T}]$ (blue-dashed) vs. normalized frequency detuning $(\delta+\omega_{m})/\gamma_{m}$ with $\omega_{m}=\kappa_{1}/4=\kappa_{2}=10^{4}\gamma_{m}$ under the conditions in Eqs. (16) and (17).
It can be seen from Fig. 6 that with these parameters, the transparency window (red-solid) appears at detuning $\delta=-\omega_{m}$, accompanied by a positive dispersion slope (blue-dashed) which is entirely different from the above OMIT at $\delta=\omega_{m}$. In fact, it is just the positive dispersion slope that causes the fast light effect in the system.

According to Eq. (5) and with the conditions in Eqs. (16) and (17), the dispersion curve slope at the transparency window ($\delta=-\omega_{m}$) can be obtained as 
\begin{eqnarray}
\mathrm{K}_{\mathrm{max}}=\frac{4\kappa_{1}[\kappa_{2}(\kappa_{2}^{2}+4\omega_{m}^{2}-\kappa_{1}\kappa_{2})+2\gamma_{m}\omega_{m}^{2}]}{\gamma_{m}(\kappa_{2}^{2}+4\omega_{m}^{2}-\kappa_{1}\kappa_{2})^{2}}.
\end{eqnarray}
Similarly, according to Eqs. (5) and (12), we obtain the maximum time delay at detuning $\delta=-\omega_{m}$ as
\begin{eqnarray}
\tau_{\mathrm{max}}=-\frac{4\kappa_{1}[\kappa_{2}(\kappa_{2}^{2}+4\omega_{m}^{2}-\kappa_{1}\kappa_{2})+2\gamma_{m}\omega_{m}^{2}]}{\gamma_{m}(\kappa_{2}^{2}+4\omega_{m}^{2}-\kappa_{1}\kappa_{2})^{2}},
\end{eqnarray}
which is exactly equal to $-\mathrm{K}_{\mathrm{max}}$ in Eq. (18). Here, $\tau_{\mathrm{max}}$ is always negative (due to $J^{2}>\kappa_{1}\kappa_{2}$), which corresponds to fast light effect. 
In Fig. 7, we plot the normalized dispersion curve slope
$\gamma_{m}\mathrm{K}$ (blue-dashed) and time delay $\gamma_{m}\tau$ (red-solid) vs. normalized frequency detuning $(\delta+\omega_{m})/\gamma_{m}$ with parameters $\omega_{m}=\kappa_{1}/4=\kappa_{2}=10^{4}\gamma_{m}$. It can be seen from Fig. 7 that the dispersion curve slope $\mathrm{K}$ and time delay $\tau$ will take the
maximum value at the transparent window $\delta=-\omega_{m}$. The maximum values $\gamma_{m}\mathrm{K}_{\mathrm{max}}=16.0032=-\gamma_{m}\tau_{\mathrm{max}}$ according to Eq. (18) and (19), which are consistent with the results in Fig. 7.

\begin{figure}[ptb]
	\includegraphics[width=0.45\textwidth]{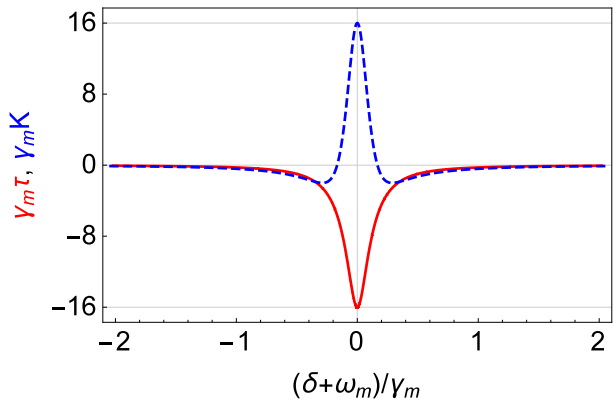}\caption{The normalized time delay $\gamma_{m}\tau$ (red-solid) and the dispersion curve slope
		$\gamma_{m}\mathrm{K}$ (blue-dashed) vs. normalized frequency detuning $(\delta+\omega_{m})/\gamma_{m}$ are plotted with the same parameters in Fig. 6.}%
	\label{Fig7}%
\end{figure}

\begin{figure}[ptb]
	\includegraphics[width=0.45\textwidth]{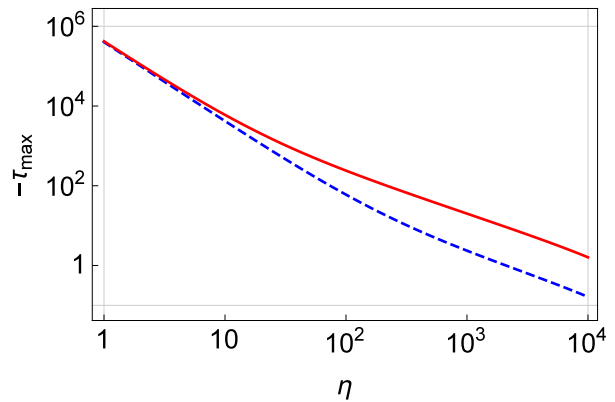}\caption{The time delay $-\tau_{\mathrm{max}}$ vs. the difference $\eta$ ($=J^{2}/\kappa_{2}-\kappa_{1}$) is plotted according to Eq. (19) with parameters $\omega_{m}=\kappa_{2}=10^{4}$ and $\gamma_{m}=10$ (red-line), $\gamma_{m}=100$ (blue-dashed) under the conditions in Eqs. (16) and (17).}%
	\label{Fig8}%
\end{figure}

It can be seen from Eq. (19) that if $\kappa_{1}\ll J^{2}/\kappa_{2}$ resulting that the $\gamma_{m}$ term in the numerator can be ignored, then $\tau_{\mathrm{max}}$ can be simplified as 
\begin{eqnarray}
\tau_{\mathrm{max}}=-\frac{4\kappa_{1}\kappa_{2}}{\gamma_{m}(\kappa_{2}^{2}+4\omega_{m}^{2}-\kappa_{1}\kappa_{2})}.
\end{eqnarray}
Similarly, if $\kappa_{1}\rightarrow J^{2}/\kappa_{2}$, the $\tau_{\mathrm{max}}$ in Eq. (19) can be simplified as 
\begin{eqnarray}
\tau_{\mathrm{max}}=-\frac{8\kappa_{1}\omega_{m}^{2}}{(\kappa_{2}^{2}+4\omega_{m}^{2}-\kappa_{1}\kappa_{2})^{2}},
\end{eqnarray}
which means that in this case the $\tau_{\mathrm{max}}$ will not be related to $\gamma_{m}$. 
In Fig. 8, we plot the time delay $-\tau_{\mathrm{max}}$ vs. $\eta$ ($=J^{2}/\kappa_{2}-\kappa_{1}$) according to Eq. (19) with $\omega_{m}=\kappa_{2}=10^{4}$ and $\gamma_{m}=10$ (red-line), $\gamma_{m}=100$ (blue-dashed). It can be seen from Fig. 8 that the ultrafast light can be achieved with small $\eta$ ($\kappa_{1}\rightarrow J^{2}/\kappa_{2}$), and in this case, all curves with different mechanical decay rates $\gamma_{m}$ will tend to the same value given by Eq. (21), while the $\tau_{\mathrm{max}}$ will tend to the value given by Eq. (20) for large $\eta$.

\section{Conclusion}

In summary, we have theoretically studied the optomechanically induced ultraslow and ultrafast light in a passive-active optomechanical system based on the ideal optomechanically induced transparency (OMIT). We first obtain the conditions of the ideal OMIT, and under which we attain the analytic expressions about the dispersion curve slope and the
optical group delay of the transmitted light. From the theoretical results, we can draw some important conclusions: (1) The optical group delay is exactly equal to the negative value of the dispersion curve slope at the ideal transparency window ($\delta=\pm\omega_{m}$). (2) at the transparency window $\delta=\omega_{m}$, the ultraslow light can be easily achieved by adjusting the difference between the decay rate $\kappa_{1}$ and the gain rate $\kappa_{2}$, even with the usual mechanical linewidth (such as Hz linewidth), and the time delay can easily exceed the mechanical ringdown time (the upper bound of the time delay in one-cavity optomechanical system); (3) similarly, at the transparency window $\delta=-\omega_{m}$, the ultrafast light can be achieved by adjusting the coupling strength and the dissipation rates in the system. We believe these results can be used to control optical transmission in quantum information processing.

\end{document}